\documentstyle[prl,aps,epsf,multicol]{revtex} 

\begin{document}
\draft
\author{D. van Effenterre~\footnote{Corresponding author. Email address:
d.van-effenterre@college-de-france.fr}, R. Ober,
 M.P. Valignat, and A.M. Cazabat}
\address{Physique de la Mati{\`e}re Condens{\'e}e, Coll{\`e}ge de France, URA 792 du CNRS,
 11 place Marcelin Berthelot, 75005 PARIS, France}
\title{Binary separation in very thin nematic films: thickness and phase coexistence}
\date{April 26, 2001}
\maketitle
\begin{abstract}
  The  behavior as a function of temperature of
 very thin films (10 to 200 nm) of pentylcyanobiphenyl (5CB) on silicon substrates is reported.
  In the vicinity of the nematic/isotropic transition we observe a 
  coexistence of two regions of different thicknesses: thick 
  regions are in the nematic state while thin ones are in the 
  isotropic state. Moreover, the transition temperature is shifted 
  downward following a $1/h^2$ law ($h$ is the film thickness). Microscope observations and small angle X-ray scattering  
allowed us to draw a phase diagram which is explained in terms of a binary first order phase
transition where thickness plays the role of an order parameter.

\end{abstract}


\begin{multicols}{2} 

\par The effect of confined geometry on the phase transitions of liquid crystals has
 attracted much attention over the years, from both
theoreticians and experimentalists \cite{SHE76,SLU90,KUZ83,IAN92,BEL92,GOL88,YOK88,WIT96}. In 
particular Sheng \cite{SHE76} first predicted a shift in the 
nematic to isotropic transition temperature for a thin nematic 
film held between two ordering surfaces and a change from first 
order to second order transition under a critical thickness. This 
theoretical work has since been extended by Sluckin and 
Poniewierski \cite{SLU90} to the case of disordering surfaces and 
to situations where the sandwiching surfaces present 
either identical or competing orientations. 
 In the meantime, experimental studies on liquid
crystals in porous media \cite{KUZ83,IAN92,BEL92}, in
submicron-sized droplets \cite{GOL88} or  in thin
films by Yokoyama \cite{YOK88} and more recently by Wittebrood {\it et al.} \cite{WIT96}
have shown confinement induced shifts in the
transition temperature consistent with theoretical
predictions. Yet in most experimental cases, no conclusion could be drawn firmly on the order of the 
transition. Liquid crystalline films have also 
attracted much attention recently in dewetting studies ~\cite{HER98,VAN99,ZIH00,OUR99}.
                             
\par   In this letter, we study the case of very thin films, the thickness $h$ ranging from 
10 to 200 nm, and report the first experimental evidence of
a coexistence between two phases at equilibrium corresponding to two different thicknesses.
 This possibility of 
the system to adapt the thickness, as one boundary is a free surface,
 has neither been observed previously nor
 been predicted theoretically, yet it is a very striking effect of the surface on phase 
 transitions. 
 The decrease in the transition
temperature shows a $1/h^2$ dependence consistent with Sluckin predictions, but 
the system adopts a binary state, here interpreted as a first 
order phase transition coupled with an elastic distortion field.
This observation may be relevant to explain experimental data of some other groups  as it shows
that an important parameter -the thickness- may, in certain cases, not be fixed by the experiment.

\par For these experiments we use 4-n-pentyl-4'-cyanobiphenyl (5CB)
(purchased from BDH Ltd., purity 99.5 \%), a liquid
crystal that undergoes a bulk phase transition from the nematic
 to the isotropic state at $35 ^{\circ}{\rm C}$. The solid substrates we use are
silicon
  wafers (type $n$, dopant $P$, purchased from Siltronix), bearing a 
  natural,
$2\ {\rm nm}$ thick, amorphous silica layer. They are used with no further cleaning or chemical treatment. The
substrate roughness measured by X-ray reflectivity is $.35\ {\rm
nm}$. 
 The anchoring of 5CB molecules on the substrate is known to
be planar (i-e parallel to the surface) \cite{VAN98} while the anchoring at the air
is homeotropic (perpendicular to the surface). 
 The homogeneous film is obtained by spin-coating a
solution of 5CB diluted in ultra-pure chloroform and the thickness
is controlled by adjusting either the concentration of the
solution or the velocity of the spin-coater (typically $3000\ {\rm
rpm}$). All the films are formed at room temperature ($23
 ^{\circ}{\rm C}$), where the 5CB is in the nematic phase in the bulk. The
thickness is then measured very accurately either by Small Angle
X-Ray Scattering (SAXS) or ellipsometry while the film is being
observed under a reflection microscope, either directly or between
crossed-polarizers, the image being acquired and analyzed on a
computer. The experimental set-up includes a temperature
controller able to maintain the temperature and its homogeneity in the film
 to better than $.05
^{\circ}{\rm C}$.

\par After forming a film of a given initial thickness $h$ ranging
from $\sim10 \ {\rm nm}$ to $\sim200 \ {\rm nm}$, we observe its
behavior when heated from nematic phase or cooled from isotropic
phase. For each chosen temperature, we wait a few minutes for the
system to reach equilibrium, the state then being stable over
hours. A representative example is given on Fig.1 for a film of initial thickness $h =48.5
\ {\rm nm}$, with the associated characteristic temperatures (the
dark regions here are higher in thickness than the clear ones).
For this thickness, when the temperature is lower than $T_1=33.1 ^{\circ}{\rm C}$ or
higher than $ T_2=34.5 ^{\circ}{\rm C}$, the film is homogeneous and
presents the same thickness as when formed at room temperature.
For any temperature between $33.1  ^{\circ}{\rm C}$ and $34.5
^{\circ}{\rm C}$, the film presents coexisting regions of two different
thicknesses.
On a short time scale, coalescence modifies the shape and distribution 
of these regions before stabilization but the thicknesses are not affected. On a much longer
time scale (few hours), the system evolves because
of a slow dewetting.
The temperatures have been given here for this
particular film, but both $T_1$ and $T_2$  depend on the initial
thickness.


\begin{figure}
\epsfxsize=9cm
\centerline{\epsfbox{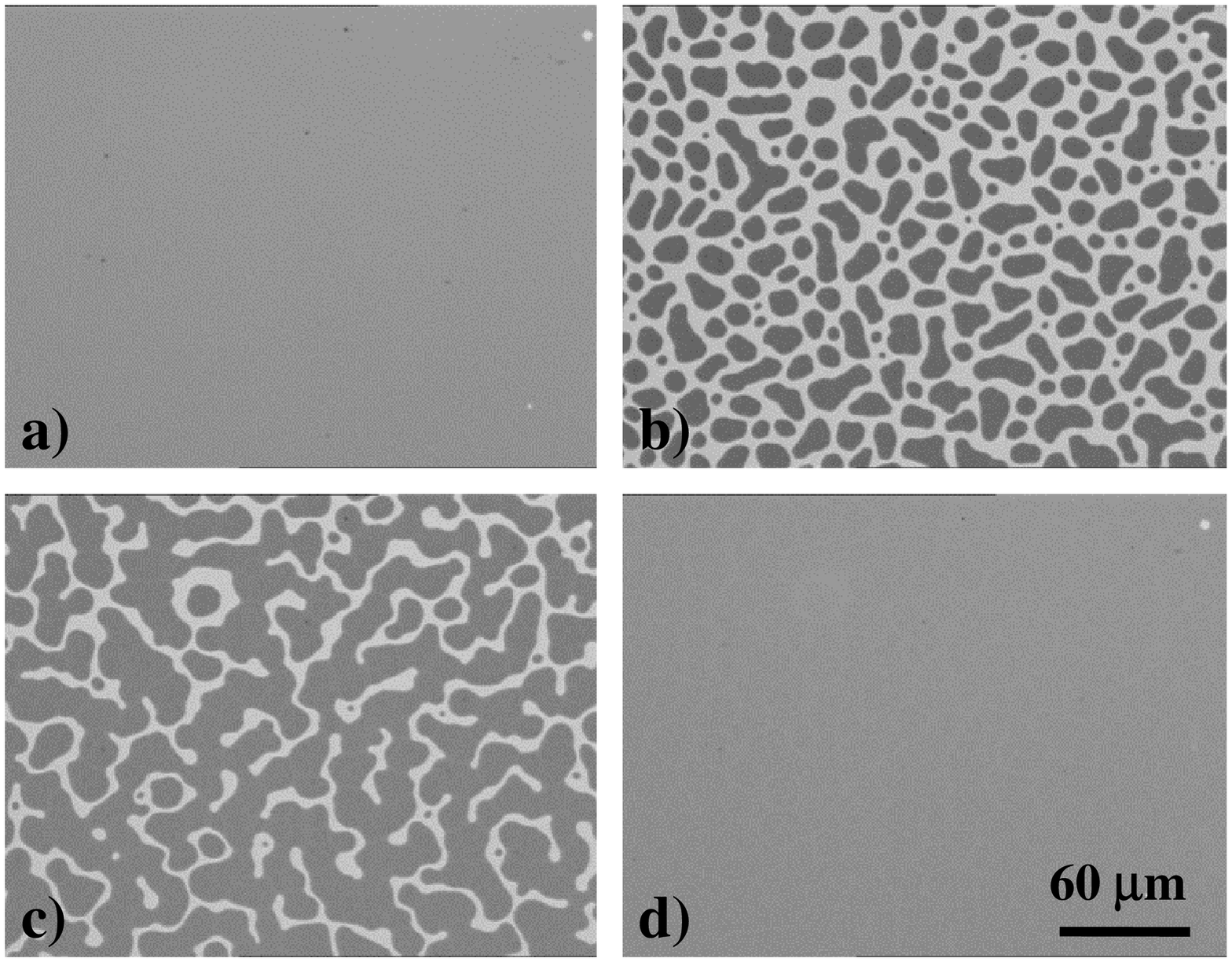}}
 {\small FIG. 1. Microscope Images
of a $48.5\ {\rm nm}$ thick film for different temperatures when
cooled down from isotropic phase: a) $T=36\ ^{\circ}{\rm C}$ ;
b) $T=33.9\ ^{\circ}{\rm C}$ ; c) $T=33.4\ ^{\circ}{\rm C}$ ; d) $T=32\
^{\circ}{\rm C}$. We notice the coexistence of two different
thicknesses for the two intermediate temperatures.} \label{nem}
\end{figure}  

\par A more detailed observation of the process leading to the
stable state when cooling the film from the isotropic phase shows
that (Fig.1): $i)$ at temperature $T_2$ -which is below the bulk
$T_{NI}$- regions of thicker film ("islands") nucleate, leading to
a situation where two different thicknesses coexist, with a
certain area fraction (defined as the ratio of thick covered area over total area);
 $ii)$ decreasing the temperature further, regions of
thicker film decrease in thickness and the area fraction increase, while the thinner regions
also decrease in thickness; $iii)$ finally the film recovers its
initial homogeneous thickness at temperature $T_1$.  For example,
we observe that for $33.9 ^{\circ}{\rm C}$ (Fig. 1b), the
thickness of the dark regions is higher ("darker") than the
thickness of the dark regions of  a lower temperature $33.4
^{\circ}{\rm C}$ (Fig.1c) but that the area fraction is smaller. We checked that
the variation of the area fraction is compatible with a volume
conservation law.

\par In the same manner, when
heating the film from the nematic state, it shows a similar
behavior but now regions of thinner film ("holes") nucleate.
 Moreover a comparison of two films of different initial thickness at the same
intermediate temperature shows that both coexisting thicknesses of
the two films are similar but that the area fraction is different,
the thinner film having a smaller fraction than the thicker one. 
Out of these qualitative observations we conclude that the two
coexisting thicknesses depend only on temperature while the area
fraction is a function of the chosen temperature and the initial
thickness.

\par We have used Small Angle X-ray  Reflectivity to characterize
thickness and roughness. Figure 2 shows the experimental data for
a film of initial thickness $42 \ {\rm nm}$, for three different
temperatures,  one below $T_1$, one intermediate, and one above
$T_2$ (respectively $25 ^{\circ}{\rm C}$, $33.4 ^{\circ}{\rm C}$
and $37  ^{\circ}{\rm C}$ in this example). For the two extreme temperatures, where the film is
homogeneous,
 we observe typical Kiessig interference fringes.

  The data are easily fitted  using Parratt's method \cite{PAR54} by a calculation of the intensity
   scattered by an homogeneous film of electronic density $\rho_e=0.33\  {\rm e\cdot\AA^{-3}} $,
    which corresponds to 5CB electronic density.
     We obtain a thickness of $42.0 \pm 0.2 \ {\rm nm}$
    with a roughness of $.50{\pm .05 \ {\rm nm}}$ for $37  ^{\circ}{\rm C}$,
     where the film is in the isotropic phase,
    and a thickness of $39.8\pm.2 \ {\rm nm}$  with a roughness of
    $.80 \pm .05 \ {\rm nm}$ for $25 ^{\circ}{\rm C}$, where the film
     is in the nematic phase (the slight difference in the thickness, typically 
     4\%, is only due to the long time necessary to acquire a sequence of X-ray spectra).

 \begin{figure}
\epsfxsize=9cm
\centerline{\epsfbox{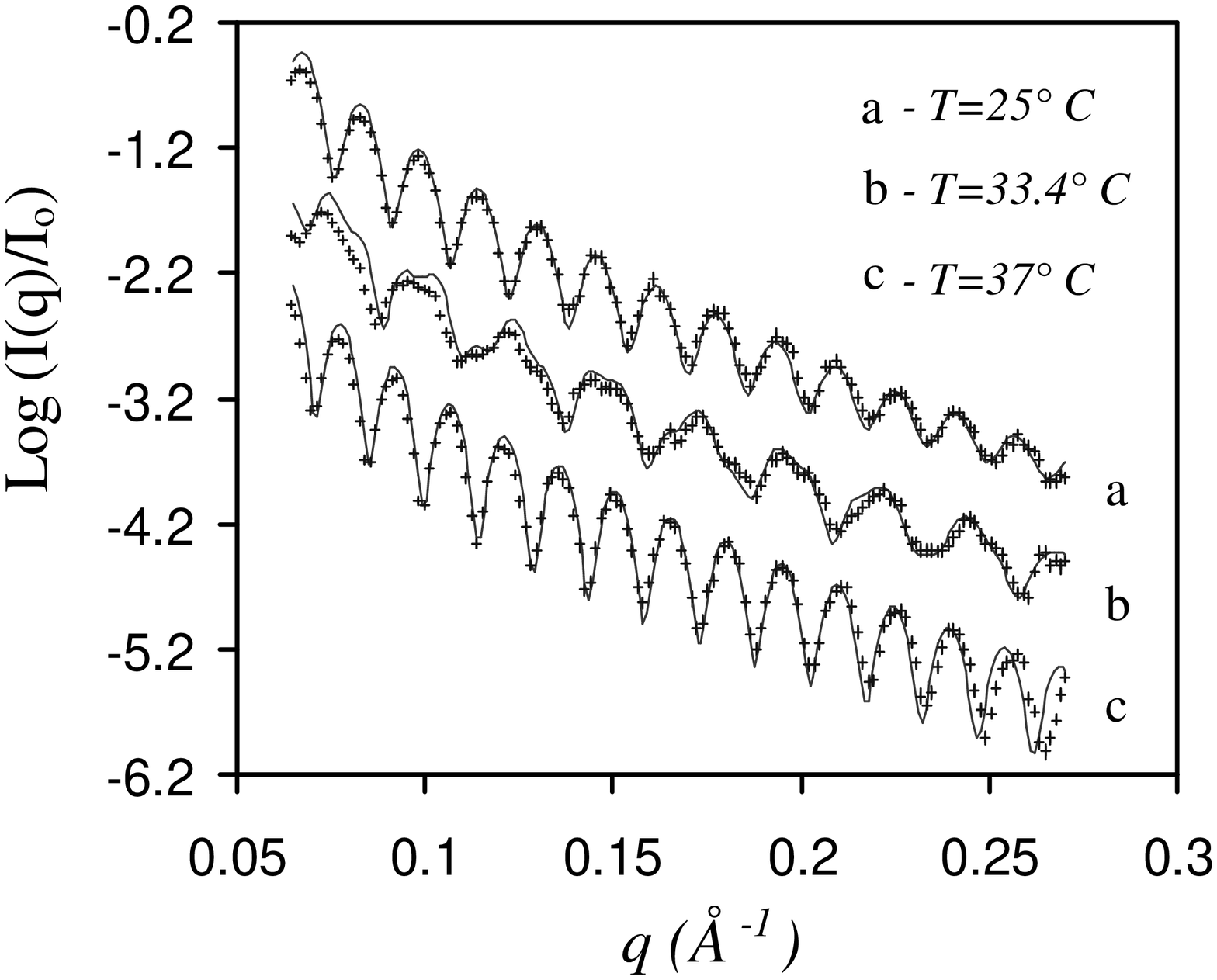}} {\small FIG. 2. Normalized X-ray
scattered intensity versus wave vector transfer $q$ for a $42\ {\rm nm}$ thick film at
three different temperatures. The lines are the corresponding
fit using Parratt's method \cite{PAR54}}. \label{rx}
\end{figure}
               
\par As for an intermediate temperature,
where the coexistence takes place, the intensity does not present
classical oscillations but is the superposition of two systems of
fringes, as can be seen on the second curve of Fig.2 ($33.4
^{\circ}{\rm C}$ in this case). We measure the area
fraction ($0.44$ in this case) on a microscope image of the film and use 
it as a fixed value in the model.
The best fit, shown on Fig.2, is then obtained with a model 
of two homogeneous films
(with same electronic density $\rho_e=0.33 \  {\rm
e\cdot\AA^{-3}})$, one of thickness $25.8\pm .2\ {\rm nm}$ and
roughness $.58\pm .05 \ {\rm nm}$, the other of thickness $62.3\pm
.5 \ {\rm nm}$ and roughness $.85\pm .05 \ {\rm nm}$, with a
relative weight corresponding to the area fraction. One can check that volume is
conserved ($0.56\times 25.8+0.44\times 62.3=41.8\simeq 42\ {\rm
nm}$).
\par In those experiments, homogeneous isotropic films ($T>T_2$) systematically show a
lower roughness (from $.45$ to $.55 \ {\rm nm}$) than homogeneous nematic 
ones ($T<T_1$)
(from $.75$ to $1.2 \ {\rm nm}$). The roughness of the film is
thus an important parameter and is characteristic of the phase.
The typical value found for an isotropic phase is in good
agreement with what we expect from the thermal fluctuations of an
isotropic liquid free interface. The roughness $\bar{\sigma}$
calculated from a capillary wave model \cite{PER89} is given by
    $\bar{\sigma}^2=\langle z^2\rangle=\frac{k_B T}{2\pi\gamma}\log\frac{Q_{max}}{Q_{min}}
       \label{rug}$
where $\gamma$ is the surface tension of the liquid and $Q_{max}$
and $Q_{min}$ are two cut-off wave vectors depending on the
system, respectively the size of the molecules and the coherence
length of the X-ray beam. Taking experimental values: $\gamma=29\
{\rm mN\cdot m^{-1}}$, $Q_{max}= 4.10^9\ {\rm m^{-1}}$ (which
corresponds to an average molecular size of $16 \ {\rm nm}$) and
$Q_{min}= 8.10^4\ {\rm m^{-1}}$ (which corresponds to a measured
coherence length of $80\ {\rm \mu m}$), one finds $\bar{\sigma}=.49 \
{\rm nm}$. The higher roughness in the case of a nematic film may
be explained by the distortion of the director field:
the degenerate planar anchoring at the substrate (i-e isotropic in the 
azimuthal direction) induces defects in the configurations of 
molecules near the lower surface that propagate through the film 
to protrude from the upper free surface, which is thus distorted.

The important point  is now that, for intermediate cases, 
the film of smaller thickness
has the characteristic roughness of the isotropic phase and the
one of higher thickness has the  characteristic roughness of the
nematic phase. Simultaneous observations using crossed-polarizers  made on  films $\gtrsim40\
{\rm nm}$ show birefringence
 in thick regions contrary to thin regions\cite{REM1}. Moreover we have checked that in thinner regions the film 
 does not present a uniform homeotropic nematic state but truly an
isotropic state, using the following method: knowing the exact
thickness by SAXS, the apparent optical refractive index of the
film was then measured by ellipsometry for different incident
angles at temperatures just above $T_2$. As in the case of the isotropic phase and
contrary to the case of the nematic phase, the apparent refractive index in this region remains
constant with respect to the angle. The convergence of those observations shows that for
intermediate temperatures, the film is composed of thin regions of
isotropic phase and thick regions of nematic phase. The
description in terms of coexistence of two films of different
thickness can now be understood as the coexistence of nematic and
isotropic phase.

\par Repeating the procedure at different temperatures for each initial film thickness,
 one is
able to build up a phase diagram, as shown in Fig.3. This diagram
gathers all the different points $(h,T)$ corresponding to a state
with coexistence of two thicknesses. It presents three parts:
the upper part where the film is homogeneously in the isotropic
phase, the lower part where the film is homogeneously in the
nematic phase and the intermediate part where there is separation
in two phases.                                         
 We notice that for
thick films ($>200 \ {\rm nm}$) both characteristic temperatures
tend towards the bulk transition temperature, as 
expected from  bulk behavior.   

\begin{figure}
\epsfxsize=9cm
\centerline{\epsfbox{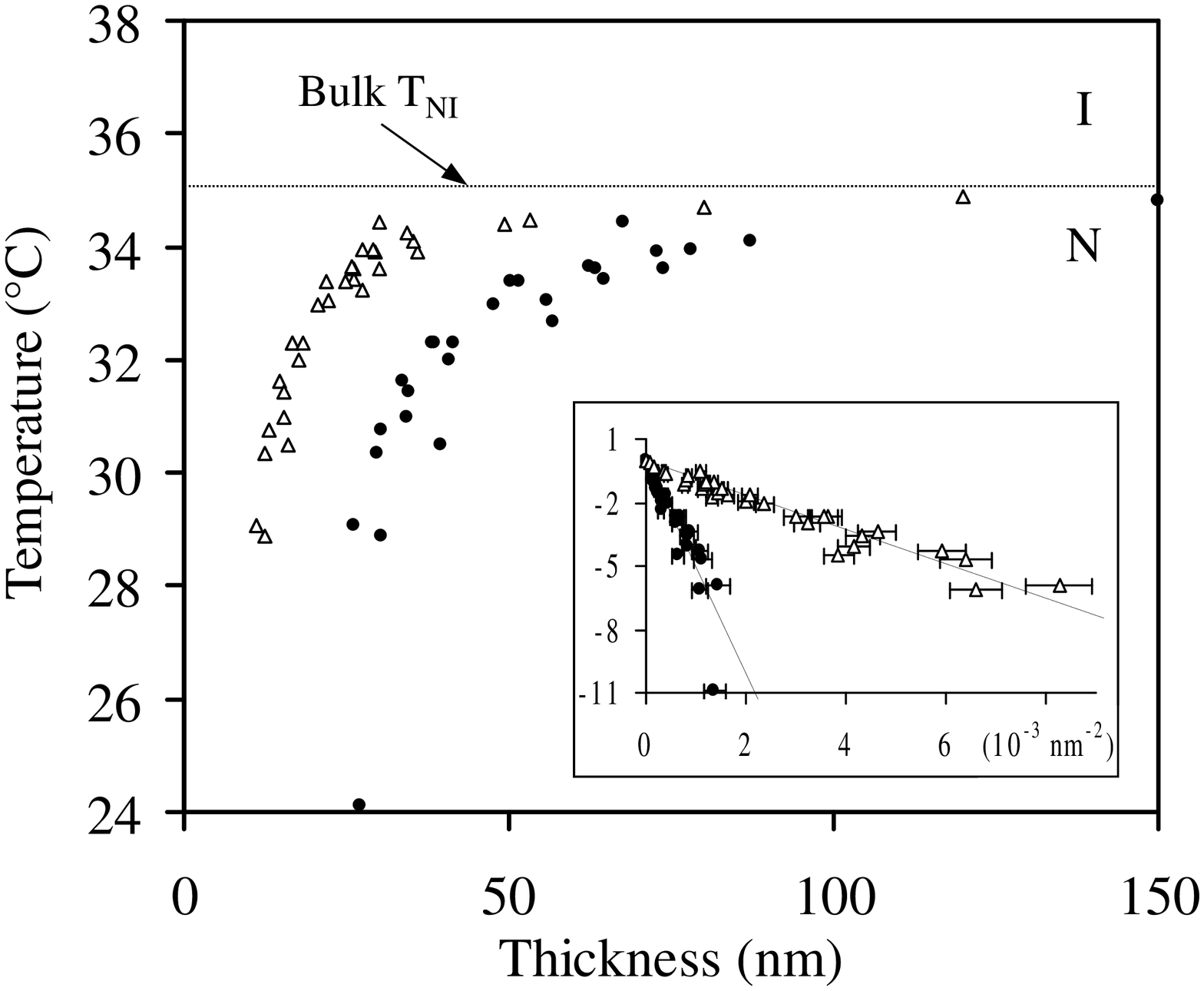}} {\small FIG. 3. Phase diagram
gathering the two coexisting thicknesses as a function of
temperature for different initial thicknesses : the open marks are
the thin isotropic phase and the filled ones the thick nematic
phase. The error bars are within the
 points dispersion and are not represented for clarity. 
 The inset presents the same data, plotting $T-T_{NI}$ as a function of $1/h^2$, with two linear fits.
  } \label{diag}
\end{figure}

\par This situation is reminiscent of the phase separation in a
binary mixture of two miscible nematogens at initial concentration $c_0$ when it
 is quenched at a temperature 
between the two pure transition temperatures: a smaller concentration $c_1$
 in the isotropic phase coexists with a higher
concentration $c_2$ in the nematic phase, with a ratio
$\varphi$ such that $c_0=\varphi\cdot c_1+(1-\varphi)\cdot c_2$.
In the same manner,
the description made above of the behavior of the film with
temperature can be followed on the phase diagram when following a
vertical line at constant $h=h_0$. The diagram has to be read
as a first order phase transition in a binary mixture, where the
thickness plays the role of the concentration, where $h_0=\varphi\cdot h_1+(1-\varphi)\cdot h_2$,
with $h_1(T)$, $h_2(T)$ and $\varphi(h_0, T)$. Here $h_1$ and
$h_2$ are also two different phases.
\par This thickness transition and the shift in the
transition temperature can be interpreted as a competition between
the energy that is necessary to melt the sample in the isotropic
state before the bulk transition temperature and the distortion
energy kept inside the film. Indeed the two antagonists anchoring 
conditions, planar and homoetropic,
imply that the director field undergoes an elastic distortion along the normal to the substrate.
Assuming there is no anchoring transition, that is to say no
 change as a function of temperature or thickness in the anchoring angles near both surfaces (an assumption
  that will be discussed further on), the
elastic component of the free energy per unit surface of the film
may be written as \cite{PGG,PRO78}:
\begin{equation}
    F_{el}(h)=\frac{1}{2}\frac{K{\Delta\theta}^2}{h}
       \label{nrjelas}
\end{equation} 
where $K$ is the elastic constant of the nematic (in the one
constant approximation) and $\Delta\theta$ is the difference between
the anchoring angles at the two boundaries.

Following Sluckin {\it et al.}\cite{SLU90}, 
a balance between this energy and the cost of moving the transition to a temperature  
$T_{NI}+\Delta T_{NI}$
 gives the Kelvin equation,
 which describes the variation of the transition temperature with thickness: 
\begin{equation}
  \frac{\Delta T_{NI}}{T_{NI}}
    =-\frac{1}{2\Delta H_{NI}}\cdot\frac{K{\Delta\theta}^2}{h^2}
       \label{shift}
\end{equation}
where $\Delta H_{NI}$ is the latent heat of the transition. 
 The experimental 
$-1/h^2$ dependence of both coexisting lines of the transition is shown 
in the inset of Fig.3, where the fits give 
$T-T_{NI}=-\frac{C}{h^2}$, with $C=5.04\ 10^{-15}\ {\rm K.m^{2}}$ 
for $T_1$ and $C=0.82\ 10^{-15}\ {\rm K.m^{2}}$ 
for $T_2$ respectively. Using
experimental data given in the literature for 5CB: $\Delta
H_{NI}\sim10^6\ {\rm J.m^{-3}}$\cite{YOK88}, $K\sim3\times10^{-12}\ {\rm N}$\cite{BRA85} and
 $T_{NI}\sim300\ {\rm K}$ and $\Delta\theta\sim\frac{\pi}{2}$, we estimate $C\sim10^{-15}
\ {\rm K.m^{2}}$, which is of the correct order of magnitude. The good quantitative agreement
 between experiments and these theoretical predictions must not hide
  the fact that the existence of two coexisting thickness is not 
  explained simply by the Kelvin equation: the thickness separation can only be explained
   once conservation of volume is taken into account. A homogeneous film will have to pay the price of the latent 
heat to be isotropic or the price of the elastic distortion to be nematic. 
  For a certain range of temperatures depending on the initial 
  thickness,
 it can be energetically advantageous for the system to 
separate into two thicknesses: the sum of the energy of a nematic higher regions -favorable because releasing the 
elastic distortion- and the energy of an isotropic lower regions -unfavorable  because isotropic- may be
 lower than the energy
 of an homogeneous film, either in the isotropic state or in the nematic state at this temperature.  A model within a Landau-de-Gennes
framework confirms quantitatively this assumption  and will be detailed in a forthcoming paper \cite{VAN01}.
This also accounts for the fact that no anchoring transition occurs in this system, 
as this scenario will allow the system to keep any nematic thickness  
above the extrapolation length $K/W$ (where $W$ is the anchoring strength), which  describes 
the critical thickness below which a constrained nematic will prefer 
to break the anchoring rather than undergo the elastic distortion.
This interpretation shows that the thickness plays the role of an 
order parameter. This reveals the first order nature of the phase transition where a 
binary state can be observed. 
    
\par In conclusion, we have shown in this letter that for very thin
cyano-biphenyl films the nematic to isotropic phase transition is
coupled with a thickness transition, leading to intermediate
states where thin regions of isotropic phase coexist with thick
regions of nematic phase. We
have been able to build experimentally  a phase diagram describing
this transition and showing the dependence of the transition
temperature with the thickness. We have interpreted this behavior
as the result of a competition between the elastic energy and the
latent heat of the transition, coupled with a volume conservation law.
 We have also observed this behavior
for compounds of the same family (nCB) and it is certainly more
general. It may play a significant role in the stability of
LC films, in the sense that we can not impose an arbitrary thickness at any
temperature for such systems. Moreover it may also mean that in such systems and in certain cases,
 a continuous variation of a measured 
parameter (light scattering, birefringence 
\ldots) with temperature   may not be the signature of second order phase transition 
but the result of mixing of two signals coming from two coexisting regions in 
a different state, during a first order phase transition.  Being aware of this behavior may be
relevant for the fundamental or theoretical study  of these
systems.
\par We would like to greatly acknowledge Didier Roux and Jean Daillant
 for fruitful discussions.

%
%
%
\end{multicols} 


\begin{references}
\bibitem{REM1} for$\lesssim40\ {\rm nm}$ films, birefringence is too 
weak.
\end{references}
\end{document}